\documentclass[12pt]{article}
\usepackage{fullpage}
\usepackage{authblk}

\usepackage{times}  
\usepackage{helvet} 
\usepackage{courier}  
\usepackage[hyphens]{url}  
\usepackage{graphicx} 
\usepackage{graphicx}  
\usepackage{amsthm}
\usepackage{amssymb}
\usepackage{mathrsfs}
\usepackage{amsmath}

\usepackage{caption}
\usepackage{subcaption}
\usepackage{graphicx}
\usepackage{color}

\usepackage{amsmath,amssymb,mathrsfs,mathtools,amsfonts,amsthm}
\usepackage[square,sort,comma,numbers]{natbib}
\allowdisplaybreaks

\usepackage{comment,color}

\usepackage{algpseudocode}

\begin{document}


\title{Peer-to-Peer Energy Trading in a Microgrid Leveraged by Smart Contracts}


\author{Guilherme Vieira, \,\ Jie Zhang\footnote{Corresponding author.}}
\affil{University of Southampton, U.K.\\ 
gbv1e17@soton.ac.uk, \,\  jie.zhang@soton.ac.uk}
\date{}


\maketitle
\begin{abstract}
The current electricity networks were not initially designed for the high integration of variable generation technologies. 
They suffer significant losses due to the combustion of fossil fuels, the long-distance transmission, and distribution of the power to the network.
Recently, \emph{prosumers}, both consumers and producers, emerge with the increasing affordability to invest in domestic solar systems. Prosumers may trade within their communities to better manage their demand and supply as well as providing social and economic benefits. 
In this paper, we explore the use of Blockchain technologies and auction mechanisms to facilitate autonomous peer-to-peer energy trading within microgrids. We design two frameworks that utilize the smart contract functionality in Ethereum and employ the continuous double auction and uniform-price double-sided auction mechanisms, respectively. We validate our design by conducting A/B tests to compare the performance of different frameworks on a real-world dataset. The key characteristics of the two frameworks and several cost analyses are presented for comparison. Our results demonstrate that a P2P trading platform that integrates the blockchain technologies and agent-based systems is promising to complement the current centralized energy grid. We also identify a number of limitations, alternative solutions, and directions for future work.
 \end{abstract}

\section{Introduction}
The current infrastructure of the national grid is aging and has been mainly built to support a one-way power flow. It is a centralized system that energy is generated in power plants, transmitted over long distances to consumptions sites, where consumers are majorly passive. See Figure 1(a) \citep{ElecGenerationFigure} for an illustration. The long-distance transmission not only creates power losses but also does not integrate demand and supply in real-time, which causes overproduction. For example, during winter 2017, Germany has experienced unexpected negative electricity prices, where they produce too much when there is not enough demand (over the weekends). As a consequence, they have to pay to be able to ``sell" their product. Thus, people have been looking for alternatives to mitigate these problems.
  
\begin{figure}[h]
 \begin{subfigure}[t]{0.4\textwidth}
 \hspace{-0.5em}
  \includegraphics[width=80mm,scale=0.9]{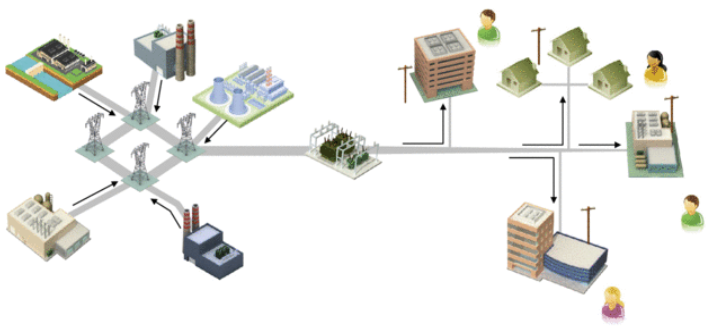} 
  \end{subfigure}\,\,\,\,\,\,\,\,\,\,\,\,\,\,\,\,\,\,\,\,\,\,\,\,\,\,\,\
  \begin{subfigure}[t]{0.4\textwidth}
  \hspace{-1em}
  \includegraphics[width=80mm,scale=0.9]{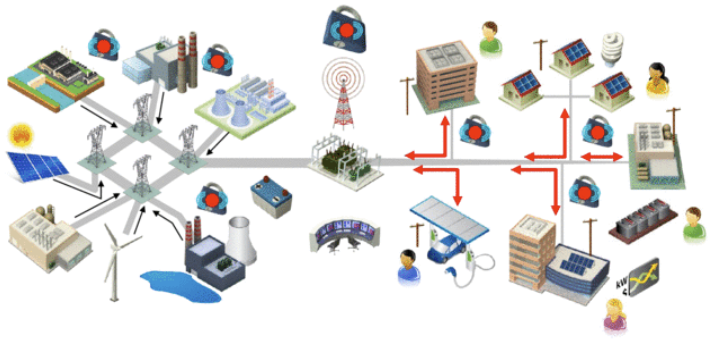}
  \end{subfigure} 
     \caption{\footnotesize (a) Central generation, one-way power flow; \,\,\,\,\,\,\ (b) Distributed generation \& storage, two-way power flow.}
    \label{fig:generation}
\end{figure}

Meanwhile, we are moving to a more distributed network, arguably, also more complex, in which energy generation will no longer only come from power plants. This distributed network involves houses with solar photovoltaic panels (PV), electric vehicle (EV) charging stations (that can act as storage), locally produced wind power, and other renewable energy technologies.
Individuals are no longer just consumers but are actively taking part in the productions of their energy and collectively feeding electricity into the grid. They are the so-called \emph{prosumers}, both consumers and producers of energy. 
That is to say, the large companies will no longer be the only ones with a stake in the energy space, but the public as well. This new movement creates a range of opportunities to retrofit the centralized way of managing today's grid, one of which is the possibility to introduce the ability to react in real-time to the intermittent generation and volatility of the national grid in the current wholesale market. 

A microgrid is formed by agglomerating small-scale prosumers; they constitute a local energy market and trade energy within their community. A microgrid can help with reflecting the real-time prices of energy and facilitate a sustainable and reliable way of locally balancing generation and consumption. The local energy markets can be viewed as a rising form of \emph{sharing economy}, in addition to the well-developed house-sharing and transportation sharing. See Figure 1(b) \citep{ElecGenerationFigure} for an illustration. Such an initiative can help reduce the latency for managing congestion and the distribution faults \citep{Jimeno2011}. Its decentralized structure can help with cyber attack resilience. Overall, it helps communities become more self-sustainable, ecologically, and economically. 

Smart grid hardware, such as smart meters, has been a significant catalyst in moving to more autonomous networks. They can be used for handling more complex information, such as optimizing energy consumption patterns, adapting to them, and using these to optimize the price of energy as well as controlling the parameters of the appliances in their houses by using the customer-generated energy profile. 
In the meantime, these systems need a secure, reliable, and transparent mechanism to manage these markets, helping with their local energy balancing needs, and providing a user-friendly application for people to use. 
Therefore, we necessitate new control frameworks to fulfill these requests.

The Blockchain technologies enable the possibility of building up an autonomous, secure, reliable, and transparent market.
Blockchain is formed by a series of blocks which contain a record of transactions. Each block references the previous block via its hash value, thereby creating a chain of transactional records that can be traced back to the genesis block. The most trusted and reliable chain is the longest. The Blockchain is not provided from a single server but is a distributed transactional database with globally distributed nodes that are linked by gossip protocols. Blockchain protocols are designed to possess a number of important features including \emph{decentralization}, \emph{immutability}, and \emph{pseudo-anonimity}. Therefore, they provide a level of security, trustless, and privacy.

{\bf Problem Statement.} 
To realize autonomous peer-to-peer energy trading within microgrids, one must demonstrate that the use of blockchain technologies to build a peer-to-peer energy trading system is technically feasible and reliable. More importantly,  it is desirable to explore associated costs and limitations.



{\bf Contributions.}
We {\bf develop two frameworks} that aggregate local demand and supply by auction mechanisms, buy and sell energy on behalf of prosumers by implementing predefined agents logic, and enable autonomous transactions through smart contracts functions. 
The {\bf evaluation} is undertaken predominantly from an economic point of view. We would compute the aggregated costs if households were to always pay to the national grid when they lack energy or not generating enough from their PV systems. This data will be compared to the results obtained from the two separate auction mechanisms. These {\bf cost analyses} would provide an insight into whether blockchain technologies are effective in integrating necessary components to constitute a micro-market setup.
Last, we reflect on the assumptions made, the challenges yet to be addressed, and point out directions for future work in these emerging applications. 

{\bf Organization.} 
This paper is organized as follows. Section 2 introduces the integral components of the system, exhibits the design of the two frameworks, and presents experimental parameters and some implementation details. Section 3 presents a high-level comparison of the key characteristics of the two frameworks and the simulation results, including some cost analyses. We conclude with a reflection on the design and discussion of future directions in Section 4. 


\subsection{Related Work}
{\bf Blockchain Technology Application in Microgrids.}
Since the emergence of Blockchain technologies, people appealed to exploit their applications in the energy sector, with the hope to reduce costs, enhance security, enable automation, and improve efficiency. This adaption requires the integration of agent-based modeling, market mechanisms, microgrid infrastructures, and many more. To the best of our knowledge, there are more attempts toward this direction in industry, and few developed prototypes documented in the literature. 
\cite{PowerSystems2019} proposed a cooperative model in which electricity producers and consumers form coalitions and collectively negotiate the electricity price. The model included a transaction settlement system such that the contracts are recorded in a Blockchain protocol. Their simulation results revealed the connection between the number of prosumers, the average negotiation time, and the number of final contracts. 
\cite{2017EI2} proposed a peer-to-peer electricity trading mechanism that enables smart contract transactions. In their mechanism, there was a single seller that aggregates local producers' supply and auction off the electricity to individual consumers. The authors ran experiments with 100 prosumers by considering their satisfaction index and exploit the relation between the demand/generation ratio and satisfaction indices. Mengelkamp et al.
\citep{MENGELKAMP2018870} contributed to the literature on a conceptual level by presenting a blockchain-based microgrid energy market and proposed a framework which contains seven market components. 
\cite{2018BigComp} investigated the pros and cons of several Blockchain protocols and concluded that Ethereum Smart Contract was most appropriate for building up a Blockchain-based electricity trading platform in Microgrid. 

{\bf P2P Energy Trading Platforms.}
There is a large body of literature on microgrids' benchmarks, benefits, and trials \citep{Mariam2013,RenewableReviews2016}. Here, we list a few projects in which the outcome was P2P energy trading platforms. 
 \cite{PowerLedger} is an energy trading platform that allows for decentralized selling and buying of renewable energy. Their platform has a dual-token ecosystem operating on two blockchain layers, POWR and Sparkz. It provides consumers with access to a variety of energy markets around the globe. 
TransActive Grid is a joint venture between a startup LO3 Energy and a blockchain tech company ConsenSys. They launched a proof-of-concept project, the Brooklyn Microgrid, which aimed at enabling local transactions based on Blockchain technologies. They did not have an Initial Coin Offer and expected to profit from transaction fees. To the best of our knowledge, they were oscillating between developing its cryptocurrencies that fuel their smart contract or employing Ethereum straightaway. There are other P2P energy trading platforms such as Grid Singularity, NRGcoin, SolarCoin, Piclo, and Vandebron.

{\bf Auction Mechanisms.}
Continuous double auction (CDA) is very common and has become one of the dominant auction models for trading financial products. The Nasdaq Stock Market and the New York Stock Exchange, for example, implement variants of the CDA. With the flourish of the electronic commerce, CDA is also adopted by many online marketplaces, such as LabX and Dallas Gold and Silver Exchange. Since \citep{DBLP:conf/ijcai/DasHKT01}, researchers have developed software-based agent mechanisms to automate double auction for stock trading with or without human interaction. More recent research on bidding in continuous double auctions and its truthfulness can be seen in \citep{DBLP:conf/ijcai/ChowdhuryKT018, DBLP:conf/aaai/Segal-HaleviHA18}.
In a uniform-priced double-sided auction \citep{DBLP:journals/sj/MaityR10,Sinha2008}, a single price at the end of each time interval is determined by the intersection of the aggregated demand and supply curve derived within the epoch. 



\section{Framework Design and Implementation}





\subsection{Components of the Model}


{\bf The Rationale of Adopting a Blockchain Protocol.} 
The most influential consensus mechanisms are Proof-of-Work (e.g., Bitcoin \citep{bitcoin2009}, Ethereum \citep{ethereum2014}, and other Altcoins) and Proof-of-Stake (e.g., Algorand \citep{DBLP:journals/iacr/GiladHMVZ17}, Ouroboros \citep{DBLP:conf/crypto/KiayiasRDO17}, Ouroboros Praos \citep{DBLP:conf/eurocrypt/DavidGKR18}). Since finding a nonce to validate transactions and create new blocks is computationally intensive, PoW is often criticized by wasting electricity. Therefore, PoS is more preferred for environmental consideration but is awaiting validation. For example, Ethereum was the first to call for a switch from PoW to PoS and proposed the Slasher and Casper protocols, and they had even set up deadlines for such a switch; however, upon till now, they are still using a PoW protocol. 

Depends on the range of nodes that can write and read the data on a chain, blockchain protocols can be classified as permissionless/public blockchain and permissioned/private Blockchain. A public blockchain (e.g., Bitcoin and Ethereum) offers a high level of decentralization, security, and immutability. Permissioned Blockchain (e.g., Hyperledger) offers a higher level of efficiency, but it is similar to a distributed system, which is challenged by fault detection, fault tolerance, and fault resilience. 

The three leading contenders that enabled smart contracts are Ethereum, Hyperledger Fabric, and NEO. Hyperledger Fabric offers an enterprise-grade framework. However, only a small portion and fragment tools are open access. Implementing a complex and integrated P2P energy trading platform would require much inferring and debugging their tools.
NEO provides fewer resources and documentation than Ethereum. Ethereum can execute relative complex business logic in a secure network. Its smart contract is a piece of code that can contain data, its balance of ether, and functions that can be triggered if the conditions/terms are fulfilled within it. Smart contracts enable a more vast range of applications that can be built on top of the Ethereum network where they can interact with each other. Therefore, Ethereum was the chosen blockchain technology.

\textbf{Ethereum Blockchain Costs.}
In Ethereum, any transaction that changes state in the blockchain database needs to be mined. Transactions that change state infer a cost, which is paid in \emph{gas}, the unit of computation in the Ethereum blockchain. Each operation (e.g., adding, multiplying) have a pre-defined cost of gas \citep{ethereum2019}. Therefore, the overall gas necessary to complete a transaction amounts to the number of operations it needs to perform up to its completion. The total cost of a transaction is given by:
\begin{equation}
\hspace{10em}
\textit{Cost of Transaction} = \textit{Gas Limit} \times \textit{Gas Price}
\end{equation}
\emph{Gas limit} is a parameter that is defined in any transaction; it aims to set a maximum limit of gas in that the user is willing to pay for said transaction. Since the simulation will be run in a local network using Ganache-cli, we set the gas price to be $2e^{10}$ Wei, where Wei is equivalent to $e^{-18}$ Ether and is the smallest unit of Ether.
One Ether will be considered to cost 250 dollars for the entirety of the analysis. Although the volatility of cryptocurrencies is beyond the scope of this work, all these parameters can be customized in the simulation file. Lastly, each transaction between user accounts has a set fee of 21,000 gas. 


{\bf Market Mechanism.} 
The market mechanism defines the conditions and rules in which energy is trade. The two mainstream market mechanisms identified in the literature, the continuous double auction \citep{MENGELKAMP2018870} and the uniform-price double-sided auction \citep{MaityRao2010,Sinha2008}, are investigated in this work. 
The continuous double auction is a mechanism often used in financial market exchanges. There are two sides, the buyer (bids) and sellers (asks) side. Each participant bids their value and quantity of the desired good; if any bid surpasses the lowest-priced ask, they are matched and trade. The uniform-priced double-sided auction \citep{DBLP:journals/sj/MaityR10} is more common in day-ahead energy markets. These markets intend to buy and sell energy for defined time intervals for the following day. This process prevents volatility in prices of energy, along with providing the reliability of energy available for the buyer. It also serves for producers to better schedule their generation capacity. The auction works in discrete intervals, in our case, set to be one hour, where all the bidders and sellers submit offers. At the end of this hour, a single price is defined by the intersection between the aggregated demand and supply curve. Trade occurs between all the participants at the equilibrium price, and the market is reset for the next hour. 



\textbf{Dataset Acquisition.}
The dataset is one of the essential components to validate our design. We acquire a dataset through the Pecan Street website, a research network with data of over a thousand homes, out of which 250 were households with installed solar systems \citep{pecan}. 
This dataset was comprised of houses mainly located in Austin, Texas, USA, providing plenty of solar coverage and with information available about their energy profile, load, demand, how much was required from the grid, and which utilities constituted the total demand.
Each of these agent instances represents a household, in which all its data (e.g., amount of battery charge, load, demand, ether balance, historical bids/asks) is stored, along with functions representing the functionality that the agent is able to perform.


{\bf Formulate the Price of Energy.} 
The data was preprocessed using Python to isolate the households into separate CSV files and calculating their respective \emph{levelised cost of energy} (LCOE), which is the sum of the system cost over a year (considering system life cycle to be 25 years) over the sum of generation that system produced - giving a \$/kWh metric. 
\begin{equation}
\hspace{10em}
\textit{LCOE} = \frac{\textit{Sum of the system cost over a year}}{\textit{Sum of generation that system produced}}
\end{equation}

The cost of the system was derived by taking the national average of price per kWh of installed solar capacity applied to the different tiers of installed capacity - e.g., 0-4 kWp, 4-10 kWp, and 10-50 kWp. Additionally, battery prices were included in the derivation of LCOE, by introducing an 8 kWh battery (Powervault Lithium-ion G200) cost for every system that had a capacity between 2 and 4 kWp, one or two Tesla power wall batteries for systems with a capacity of 4 to 6 kWp or 6 kwp and above respectively. At the end of this process, each household ended up having their two individual LCOE derived for a system without a battery, and another one with. Finally, a large biomass producer was introduced as it is a controllable renewable energy system that can be used to generate a more stable supply than solar - LCOE prices ranging from 5 to 12 cents per kWh to emulate the different types of biomass used. This provides the microgrid with self-sustainability by increasing its generation capability. 

The pricing of energy is set at a price that benefits both the producer and consumer. That is, pricing the energy above the sellers base costs but also setting a maximum price for the buyer, which did not exceed the national grid price. If any energy were to be sold at higher than the grid price, there would be no incentive for consumers to use the microgrid market. Hence, these two constraints were taken into consideration, and a normal distribution was attributed to each agent. The minimum price, i.e., the LCOE for each agent (base electricity cost), was set a two standard of deviations below the mean, and the national grid price at two standards of deviations above the mean - resulting in a \emph{truncated normal distribution}.

\subsection{Two Frameworks} 
In this section, we present two frameworks. The first one takes full advantage of the blockchain technologies, and the second one is less decentralized as it requires a component to be off-chain in order to alleviate the computational work from the blockchain. We will describe how data and information flow in the systems and how the application was going to interact with the blockchain.

\textbf{Framework 1 (Decentralized, Continuous Double Auction).} 
The first framework (Figure 2(a)) uses the full functionality of the blockchain. It is a server that households can connect to, a database with immutable information that people can access and write to, as well as a platform for the exchange of value. The bottom of the diagram indicates that each household, holding a smart meter, connects and relays information in real-time to its representative agent in the microgrid application. This connection is of restricted access to the smart meter that corresponds to the respective household owner and provides all the information the agent needs in order to make rational decisions on their behalf. If this is the first time an agent is connecting to the blockchain, the agents will request the creation of a household contract to the household factory through the Web3 interface technology, which enables agents to communicate to the blockchain. The household factory creates the household contracts on the agent's behalf and attributes their Ethereum address, respectively. It is to ensure consistency of the smart contract codes, i.e., no malicious code can be inserted by any participant if all have to be created by this factory. Moreover, it creates contracts with the decentralized exchange address automatically configured, and it is a useful point of access to request statistics about every participant's readable information. Household contacts now have a direct restricted channel to their agent, which provides and updates all the necessary information - serving as a repository of information. These are then allowed to submit bids/asks, delete them from the exchange, and make calls to the exchange when triggered by the agent's logic. The decentralized exchange takes all the bids/asks and orders them in ascending and descending order of price, respectively. This way, it automates the matching process of the offers and relays the appropriate transaction commands between household contracts. All these trades records are stored in the blockchain. The accountability of each entity's balance is taken care of by the consensus mechanism.

\begin{figure}[h]
 \begin{subfigure}[t]{0.4\textwidth}
 \hspace{-1em}
  \includegraphics[width=85mm,scale=1]{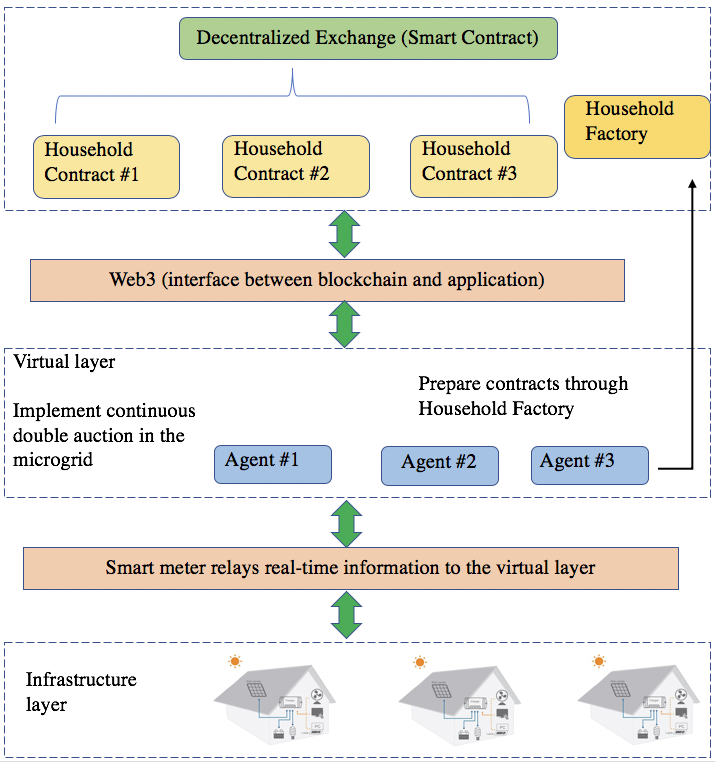}
  \end{subfigure}\,\,\,\,\,\,\,\,\,\,\,\,\,\,\,\,\,\,\,\,\,\,\,\,\,\,\,\
  \begin{subfigure}[t]{0.4\textwidth}
  \hspace{-1em}
 \includegraphics[width=85mm,scale=1]{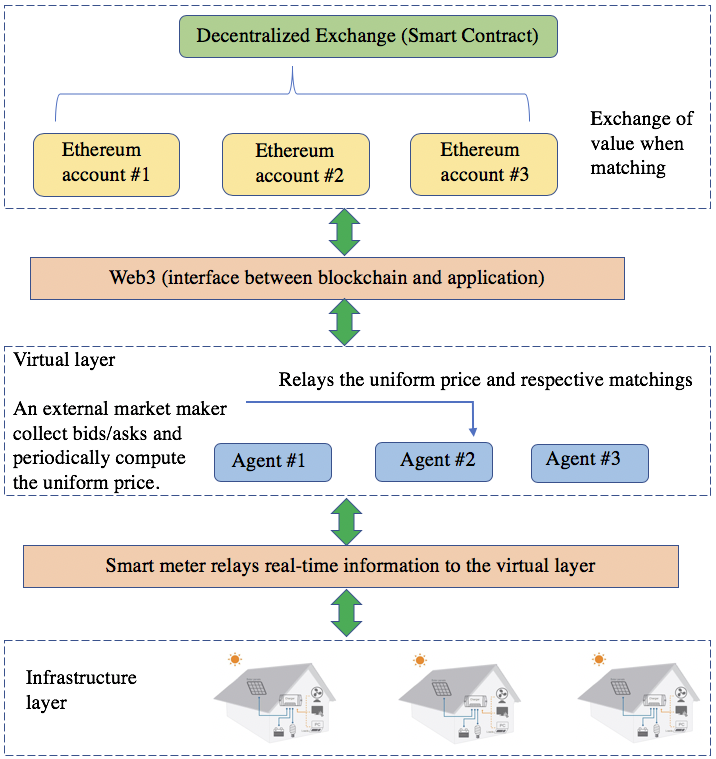}
  \end{subfigure} 
     \caption{\,\,\,\,\,\,\,\,\,\ (a) Framework 1; \,\,\,\,\,\,\,\,\,\,\,\,\,\,\,\,\,\,\,\,\,\,\,\,\,\,\,\,\,\,\,\,\,\,\,\,\,\,\,\,\,\,\,\,\,\,\,\,\,\,\,\,\,\,\,\,\,\,\,\,\,\,\,\ (b) Framework 2.}
    \label{fig:Framework}
\end{figure}

\textbf{Framework 2 (Semi-decentralized, Uniform-price Double Sided Auction).} 
The second framework is shown in Figure 2(b). Comparing to the first framework, it implements a uniform-price double-sided auction instead of a continuous double auction. More importantly, the household contacts and the factory are removed to reduce an intermediary layer of transactions that gets information onto the exchange.
The decentralized exchange no longer automates the matching of offers coming from the agents. Instead, the smart contract is used to store the information relayed from all the different participants of the microgrid. At the end of each epoch, the application collects the offers from the decentralized exchange, consequently calculating the regression curves of the supply and demand side in order to find the interception - the uniform price. The matching algorithm is then initialized, which relays the respective matching with their respective prices to each agent. The algorithm triggers their payment functionality, thereby transacting the necessary amount of Ether from one account to the other through the Web3 interface.

\begin{figure*}
  \includegraphics[width=160mm]{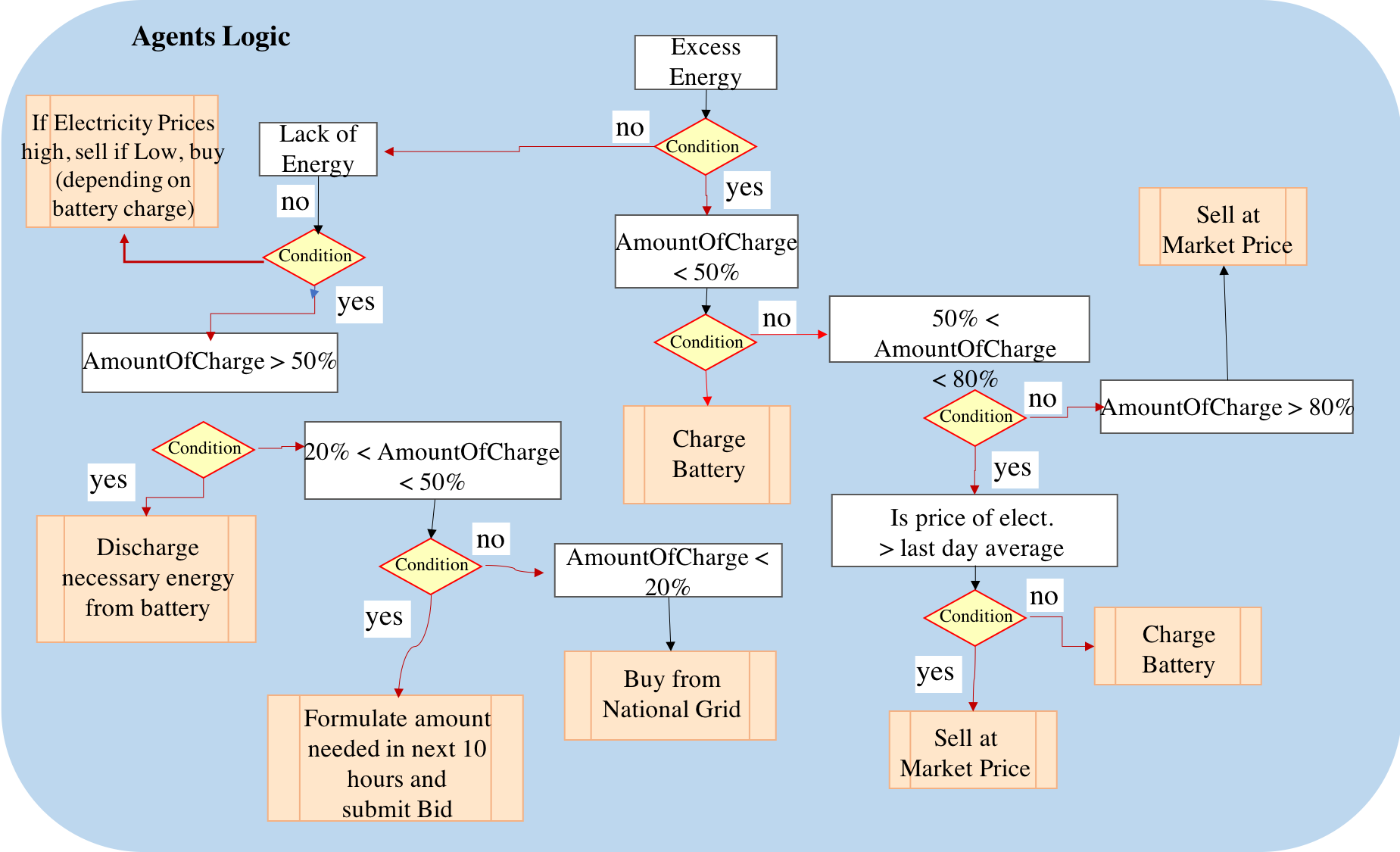}
  \caption{Digram showing the logic behind the agents strategic bidding or selling.}
  \label{fig:agentlogic}
\end{figure*}

\textbf{Agent Bidding Strategy Logic.} 
The agent bidding strategy logic is shown in Figure \ref{fig:agentlogic}. 
The first decision node from the top is checking whether an agent has excess energy or is lack of energy. Then, it follows with several thresholds of batteries in which the agent decides to sell/buy energy or to charge/discharge the batteries. This process results in a behavior in which energy is sold when prices are high and bought when prices are low. Furthermore, it has a level of "forecasting" capability in which it checks the necessary energy that is going to be needed for the following 10 hours (configurable). This attempts to emulate the actual forecasting done in energy markets in order to predict how much demand and supply individuals will need in the upcoming day. Forecasting provides a more reliable level of energy supply, as suppliers can be aware of the amount of necessary energy that will be required at different times of day, and control systems can prepare accordingly. Preprocessed data is prepared by acquiring weather data from the National Renewable Energy Laboratory (NREL) within the same time frame and location \cite{nrel}. 
This "forecasting" function permits agents to reduce the number of transactions they would have to do a day, by filling their batteries with the necessary energy they will need for that time interval. Otherwise, they would have to bid almost every hour that is inconvenient and resulting in a system that would be very costly just in transaction fees. Finally, there is a safety net mechanism in which the agents will buy energy from the national grid if the batteries fall under 20\% of their total capacity.

{\bf Implementation.}
For the implementation, a wide range of technologies are used, including Solc, Mocha, React.js, Next.js, Ganachecli, Metamask, Ganache-cli, and Web3. Ganache-cli is an Ethereum development tool that simulates full client behavior by running a personal blockchain on a local machine. It is essential to make the testing faster and easier as it can run transactions very quickly compared to the online Ethereum test networks. Web3 enables an interface to communicate between the application and the ganache provider, i.e., sending transactions, creating contracts, requesting blockchain accounts information.

The initial step taken is the creation of adequate smart contracts that would enable the storage of information and added bidding functionality to the households, name household contracts. These would then communicate to a smart contract that serves the function of a decentralized exchange market. The smart contracts are then tested using the Mocha library to check if they are working as desired and to learn about the contract's limitations - i.e., memory limitations, computational cost optimization, the functionality of transaction between contracts, and between Ethereum accounts. Once the smart contracts are sufficiently refined, the simplified UI application is developed using React.js and Next.js. 

We program the market simulation in Javascript as the underlying UI application to match the stack of technologies that are used to code the front-end of the application. We start by importing the respective CSV files containing each household's hourly load and demand into separate class instances (data structures) called agents. Each of these agent instances represents a household, in which all its data (e.g., amount of battery charge, load, demand, ether balance, historical bids/asks) is stored, along with functions representing the functionality that the agent is able to perform. These are used to interact either with its own data structures or to interact with the smart contracts. The agents are coded to take independent actions based on the most attractive conditions in the market. No coordination between the agents is built into the system. At every epoch, each agent is given the chance to act according to these sets of actions. At the end of the simulation, the data is collected from the separate agent data structures, and written to a CSV file to be examined by a Python script (better visualization and mathematical tools). 

The analysis of the results would be undertaken by calculating the aggregated costs of all the households if they were to always pay to the national grid when they lack energy or not generating enough from their PV systems. This data is then compared to the results obtained from the two frameworks. This analysis will be done predominantly from an economic point of view. These results will be discussed to assess whether using microgrids is economically beneficial for both consumers and producers, as well as providing an insight into whether blockchain technologies are effective in integrating necessary components to constitute a micro-market setup.


\section{Evaluation}
\textbf{A high-level comparison of the key characteristics of the two frameworks.} 
A comparison of the two frameworks is given in Table \ref{tab:keycharacteristics}. At a high-level, Framework 1 uses the storage, computation, and exchange of value function of the blockchain, whereas Framework 2 does the computation externally. It takes advantage of the fact that processing information is quicker and free of costs, hence vastly reducing the number of transactions used within its operation. However, this computation has to be done in a centralized server, which is prone to attacks as it is less secure than the data on a blockchain. Nevertheless, one could assume that the current centralized servers have the same level of security. This is necessary as the computation required to calculate the uniform price was too much to implement in a smart contract. In contrast, Framework 1 is more secure as all the information is within the blockchain, yet slower due to the more significant amount of transactions has to execute. In addition to this, it can only run continuous double-auction. Framework 1 is, therefore, more decentralized as every participant pays its share of computation of the decentralized exchange when placing a bid/ask.

\begin{table*}[]
\centering
\begin{tabular}{|l|l|}
\hline
Framework 1                                                                                            & Framework 2                                                                                                                \\ \hline
Computation and matching are done within \\ the blockchain through smart contracts                                                       & Matching is done externally to the blockchain                                                                           \\ \hline
Uses more transactions = more gas expenditure                                                          & Uses less transactions = less gas expenditure                                                                              \\ \hline
\begin{tabular}[c]{@{}l@{}}Less flexible - can not implement heavy \\ computation procedures\end{tabular} & \begin{tabular}[c]{@{}l@{}}More flexible - can implement more complex \\ procedures (e.g., other auction mechanisms)\end{tabular} \\ \hline
Slower system                                                                                          & Faster system                                                                                                              \\ \hline
More secure                                                                                            & Less secure                                                                                                                \\ \hline
\end{tabular}
\caption{The difference of the key characteristics of the two frameworks}
\label{tab:keycharacteristics}
\end{table*}

\smallskip

In the next, we present a number of simulation results. 

We implemented the two frameworks with real-world data on a weekly base, a monthly base, and a six-month base. Fig. 4(a) and Fig. 4(b) show the simulation results of Framework 1 and 2 in a week, respectively. 
At times in which there was more supply than demand in the microgrid, there were no buying and selling activities in the market. In this case, the prices of electricity were left unchanged from the previous one, as the flat lines represent. This is merely to allow better visualization of the prices. 

\begin{figure}[h]
 \begin{subfigure}[t]{0.4\textwidth}
 \hspace{-1em}
   \includegraphics[width=85mm,scale=1]{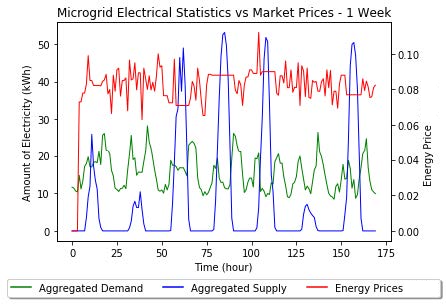}
  \end{subfigure}\,\,\,\,\,\,\,\,\,\,\,\,\,\,\,\,\,\,\,\,\,\,\,\,\,\,\,\
  \begin{subfigure}[t]{0.4\textwidth}
  \hspace{-1em}
 \includegraphics[width=85mm,scale=1]{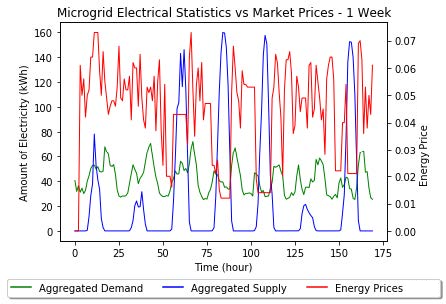}
  \end{subfigure} 
  \caption{\footnotesize (a) Framework 1, continuous double auction; \,\,\,\,\,\,\,\,\,\,\,\,\,\,\,\ (b) Framework 2, uniform-price, double-sided auction.}
\end{figure}

{\bf Electricity Prices in the Microgrid.} 
Fig. 4(a) and Fig. 4(b) show that the continuous double auction results in a higher price average of around 8.0 cents per kWh. In contrast, the uniform-price auction results in an average of 4.4 cents per kWh as well as exhibiting a more volatile range of prices. In hindsight, this is to some extent because the mechanism used in the latter does not follow individual rationality. That is, the interception between demand and supply curve may be found at lower prices than the cost of the producers (averaged at 7 cents per kWh), which results in a negative utility. The continuous double auction follows individual rationality, as consumers are paying less than they would be paying with the national grid, and producers are selling at a higher cost than their LCOE value, resulting in a beneficial situation for both parties. Furthermore, the uniform-price auction demonstrates a larger amount of electricity volume being traded. This characteristic is due to the fact that some orders do not get filled for not having a high enough price tag in relation to the sell-side in the continuous double auction.
Perhaps introducing more producers present in the market would add a stronger supply side, thereby pushing the interception of the curves to a higher price, which would lead to more reasonable prices.
Simulations of longer periods were done and are consistent with this observation. By removing the zero-activity periods, the average of prices is shown to be 8.1 cents per kWh in Framework 1 and 4.2 cents per kWh in Framework 2, in a six-month simulation.

{\bf Blockchain Framework Average Costs vs. National Grid Average Cost.}
Fig 5(a) and 5(b) demonstrate the costs comparison between using the Microgrid P2P trading platforms and the cost of using the national grid solely. The averaged total trading cost per household comes to around 0.74 dollars per day, whereas the averaged cost of using the grid per household is 2.18 dollars a day. These remain consistent with the longer periods simulation results, which show that national grid costs are around three-fold as much as the blockchain P2P trading platform.

\begin{figure}[h]
 \begin{subfigure}[t]{0.4\textwidth}
 \hspace{-1em}
   \includegraphics[width=85mm,scale=1]{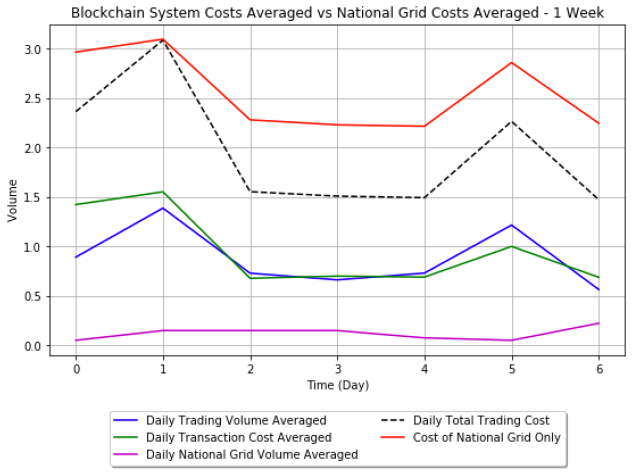}
  \end{subfigure}\,\,\,\,\,\,\,\,\,\,\,\,\,\,\,\,\,\,\,\,\,\,\,\,\,\,\,\
  \begin{subfigure}[t]{0.4\textwidth}
  \hspace{-1em}
  \includegraphics[width=85mm,scale=1]{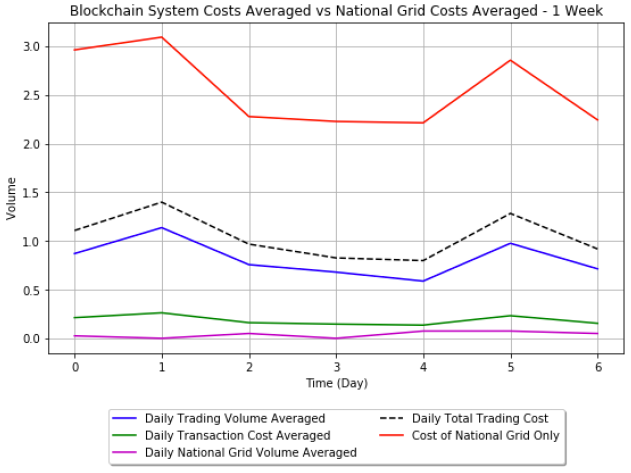}
  \end{subfigure} 
   \caption{\footnotesize (a) Average Costs of Framework 1 vs National Grid Average Cost.; \,\,\ (b) Average Costs of Framework 2 vs National Grid Average Cost.}
\end{figure}

\begin{figure}[h]
 \begin{subfigure}[t]{0.4\textwidth}
 \hspace{-1em}
   \includegraphics[width=85mm,scale=1]{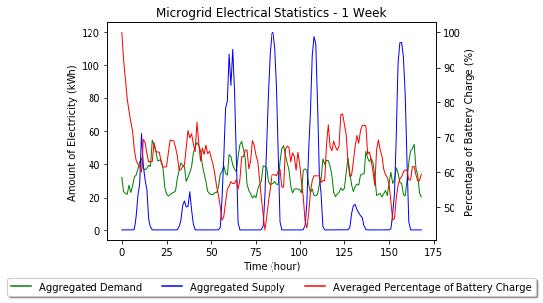}
  \label{fig:battery1}
  \end{subfigure}\,\,\,\,\,\,\,\,\,\,\,\,\,\,\,\,\,\,\,\,\,\,\,\,\,\,\,\
  \begin{subfigure}[t]{0.4\textwidth}
  \hspace{-1em}
  \includegraphics[width=85mm,scale=1]{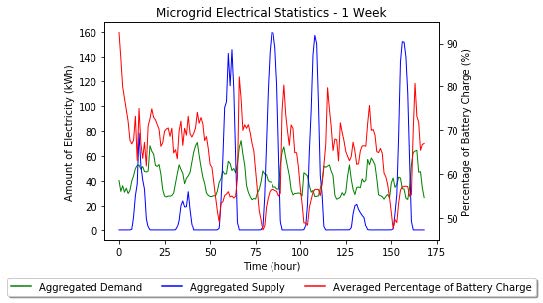}
  \end{subfigure} 
    \caption{\footnotesize (a) Average Percentage of Battery Charge in Framework 1; \,\,\,\,\,\,\,\,\,\,\,\,\ (b) Average Percentage of Battery Charge in Framework 2.}
\end{figure}

{\bf Battery Charge Percentages.}
Figure 6(a) and Figure 6(b) show the average battery charge percentages in the two frameworks, respectively. We can observe a periodic pattern in the uniform-price double auction in times when the battery charge gets to 50\%, as this triggers the Agents Logic  to purchase enough energy to cover their needs for the following 10 hours. This also coincides with the supply going up during daylight, implying that the Agents Logic correctly manages the agents' needs on a day-to-day basis. It regulates its purchases in order for the energy storage to cover its demand needs up to the following day at early hours in the morning when the sun is present when it knows it will be able to charge its battery. 
Similar patterns can be observed in the continuous double auction mechanism, where the battery charge does not tend to drop much below 50\%. The differences in volume traded in both mechanisms is reflected in the differences between these two Figures, like the spikes of battery charge in Figure \ref{fig:battery1} tend to be less abrupt, some of the bids were not filled in the order book due to their lower price. 

\section{Reflection and Future Work}
This section comprises of a critical reflection over the framework designs implemented, some results presented, along with some observations about the applicability of blockchain technologies in the microgrids use case. Some possibilities for further work are discussed.


{\bf Reflection.}
Two market mechanisms have been implemented with real-world data, showing quite different results. The uniform-price auction has shown to result in more amount of volume being traded but with a reduction in prices.

Some assumptions were made in calculating the prices of solar energy systems. Since the focus of this paper is the overall architecture design, for simplicity, it does not yet take into consideration the transmission cost, the distribution costs, and battery charging/discharging inefficiencies. 

More severely, there are still many challenges facing blockchain technologies. For example, smart contract limitations prevented the introduction of more complex logic. Also, for this specific scenario in a closed testing environment, Ethereum's transaction throughput was not a concern as this market would be trading every hour. Therefore the amount of data to the process was not a problem. As Ethereum has 15s block intervals, there could be plenty of block times to insert all the necessary transactions for an hour increment. However, this could become more of a problem as trading intervals lengthen (e.g., 1 or 5 minutes). In that case, not all transactions would be processed in time for the time interval to finalize, leaving the network congested and the market with incomplete information. 

Innovative solutions to some of these problems are under developing but would probably take some time. An example of this is the IOTA project. It is a 'ledger of things' involving the use of a tangle framework, which is computationally lighter, has low energy consumption, highly scalable, and has zero transaction fees. Indeed, this technology is promising but is a highly academic and untested framework yet, which does not have the same level of reliability as the existing blockchain protocols. Lately, the IOTA Foundation is working on additional smart contracts layer on top of their existing tangle framework, which could open new doors to a more effective, low-cost alternative solution to the problems that the blockchain faces today. It will be interesting to see how these technologies unfold in the near future and if they can fulfill their very ambitious ideas that they have set themselves.



{\bf Future Work.}
The transaction costs and efficiency of the system may be improved by optimizing the parameters in the Agents Logic algorithm. For example, the time and price at which an order is placed, and the percentage at which the battery is charged and discharged. It is essentially a minimization problem that takes into account a more substantial amount of parameters and conditions from the market, along with constraints related to the household owner's components, such as battery's depth of discharge and inefficiencies associated with them. Further testing would need more extensive simulations and a consistent framework across both market mechanisms to come to a solid conclusion about which one is most applicable. 

One may also attempt to construct alternative frameworks in which agents would communicate directly to the decentralized exchange smart contract that uses the continuous double auction (for its preferable price range and being computationally cheap). A caveat for it is that the smart contract would not be able to automate the transaction from one party to another when it found a matching. This is due to the inability of a contract to trigger transactions from a user account to another as it only has the capability to do that with contract addresses. Alternatively, a contract could be created for each successful matching/trade that would contain a balance sufficient to fill the desired amount being bought. This way, when a match is found, the market would automatically trigger the smart contract to exchange value between the two participants and return the remaining balance. Surely, any further work would have to test if the creation of said contracts for the trade would be cheaper or more costly than the current approach. 

One may want to consider agents' possible strategic behavior in designing a framework. Bidding strategy is an important problem in auction mechanisms, and many strategies for the continuous double auction have been subsequently developed. Over the last decade, there has been a considerable emphasis on strategies for software trading agents with the emergence of electronic markets \citep{Friedman1992}, such as Zero-Intelligence (ZI) strategy \citep{GodeSunder1993}, Zero-Intelligence Plus (ZIP) strategy \citep{Cliff97minimal-intelligenceagents}, GD strategy \citep{Gjerstad2001PriceFI}.  


\section{Conclusion}
The decentralized models, if fully developed, can potentially minimize the power transmission dissipation, reduce the probability of cascade failure due to its inherent islanding capability, and minimize the load on the grid by a significant ratio. The autonomous microgrid and the centralized national grid should complement each other, and contribute to a greener and economical energy market.

\setlength{\parskip}{1em}

\noindent
{\bf Acknowledgements:}
Jie Zhang was supported by a Leverhulme Trust Research Project Grant (2021 -- 2024).

\bibliographystyle{cas-model2-names}

\bibliography{blockchain}

\end{document}